# Thermalization of the Ablowitz-Ladik lattice in the presence of non-integrable perturbations

MAHMOUD A. SELIM[1], GEORGIOS G. PYRIALAKOS[2], FAN O. WU[2], ZIAD MUSSLIMANI[3], KONSTANTINOS G. MAKRIS[4,5], MERCEDEH KHAJAVIKHAN[1] AND DEMETRIOS CHRISTODOULIDES[1,*]

[1] Ming Hsieh Department of Electrical and Computer Engineering, University of Southern California, Los Angeles, California 90089, USA
[2] CREOL, College of Optics and Photonics, University of Central Florida, Orlando, Florida 32816-2700, USA
[3] Department of Mathematics, Florida State University, Tallahassee, FL 32306-4510, USA
[4] Institute of Electronic Structure and Laser, Foundation for Research and Technology-Hellas (FORTH), P.O. Box 1527, 71110 Heraklion, Greece
[5] ITCP, Department of Physics, University of Crete, 70013 Heraklion, Greece
*Corresponding author: demetri@creol.ucf.edu

We investigate the statistical mechanics of the photonic Ablowitz-Ladik lattice, the integrable version of the discrete nonlinear Schrödinger equation. In this regard, we demonstrate that in the presence of perturbations the complex response of this system can be accurately captured within the framework of optical thermodynamics. Along these lines, we shed light on the true relevance of chaos in the thermalization of the Ablowitz-Ladik system. Our results indicate that when linear and nonlinear perturbations are incorporated, this weakly nonlinear lattice will thermalize into a proper Rayleigh-Jeans distribution with a well-defined temperature and chemical potential. This result illustrates that in the supermode basis, a non-local and non-Hermitian nonlinearity can in fact properly thermalize this periodic array in the presence of two quasi-conserved quantities.

In recent years, there has been a resurgence of interest in investigating optical wave propagation phenomena in nonlinear photonic lattices. In general, such configurations can display rich light dynamics that have no counterpart in continuous settings. In this respect, optical waveguide arrays provide a fertile ground where several intriguing processes can be observed, ranging from discrete solitons [1–4] and Bloch oscillations [5,6], to dynamic localization processes [7–9] and Rabi oscillations [10]. Under tight-binding conditions, light evolution in these optical configurations can be accurately described by the discrete nonlinear Schrödinger equation (DNLS) [1-4]. Relevant to this topic, is the fully integrable version of DNLS [11], the so-called Ablowitz-Ladik (AL) model, that is known to possess an inverse scattering transform [11,12]. The integrability of this system allows one to exploit some of its hallmark features in studying, for instance, non-resonant reflectionless potentials [13], Bose-Einstein condensates [14], discrete vortices [15], dark soliton collisions [16], and optical rogue waves [17]. Moreover, the unidirectional flow of discrete solitons in a photonic lattice has been demonstrated by employing the AL potential [18]. In higher dimensions, the AL equation exhibits exotic solutions and complex waveforms, for instance, line solitons and X-shaped vortices [15]. Yet, at this point, little is known if any as to the nonlinear response of perturbed AL lattices, especially in systems where the number of modes is very large. In general, this represents a challenging problem given that the integrability of the AL model can be broken in the presence of perturbations in which case the field evolution in such a lattice becomes utterly complex and chaotic.

Quite recently, a self-consistent thermodynamic formalism has been put forward in an effort to predict in a statistical manner the response of nonlinear highly multimoded photonic systems [19–27]. This approach is universal: it can be utilized in both discrete and continuous settings in the presence of any arbitrary nonlinearities as long as more than two invariants are manifested [19]. One of the basic tenets of statistical mechanics is that the system under study is ergodic, i.e., it should be able to fully explore its phase space in a fair manner. In nonlinear multimode systems, this property naturally follows because of chaos. At this juncture, the following question naturally arises: can the AL lattice thermalize, and if so, under which conditions? It is worth emphasizing that a periodic AL lattice with $M$ sites is fully integrable given that it exhibits $M$ conservation laws. As such, it does not display chaos and thus the corresponding asymptotic Lyapunov exponents are zero [28]. Finally, in the context of the AL lattice, one may also ask whether it is indeed conceivable to identify an appropriate basis where thermalization can be explored [28].

To address this question, here we investigate the thermalization dynamics of the weakly nonlinear Ablowitz-Ladik system in both its integrable form as well as in the presence of non-integrable perturbations. In the latter case, we show that the thermalization of this lattice can be captured within the framework of optical thermodynamics [19] when the system is investigated within its linear supermode basis. When linear on-site perturbations are introduced (akin to those associated with Anderson's localization [29–31]), we show that complete thermalization into a proper Rayleigh-Jeans distribution is eventually attained at thermal equilibrium. This result illustrates that a non-local and non-Hermitian nonlinearity can in fact properly thermalize an optical lattice in the presence of two quasi-conserved quantities. Similar results are also obtained when non-integrable nonlinearities are involved.

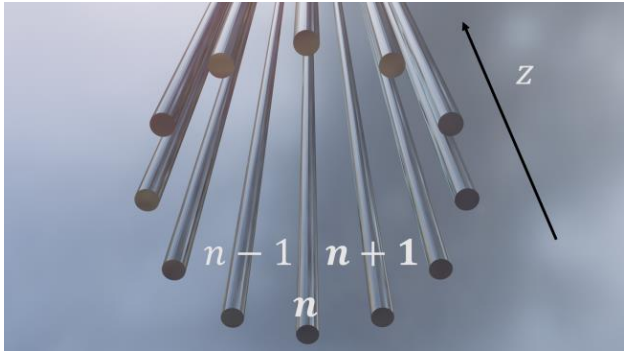

**Fig. 1.** A periodic Ablowitz-Ladik lattice consisting of $M$ coupled waveguide channels.

In general, nonlinear wave propagation in a perturbed periodic AL system consisting of $M$ waveguide sites (Fig. 1) is governed by the following evolution equation [11]:

$$i\frac{da_n}{dz} + (a_{n+1} + a_{n-1}) + |a_n|^2(a_{n+1} + a_{n-1}) \\ + \gamma |a_n|^2 a_n + \delta f_n a_n = 0, \quad (1)$$

where $a_n$ stands for the local optical field amplitude at waveguide site $n$, while $\gamma$ and $\delta$ are scaling parameters for the Kerr nonlinearity and refractive index perturbations, respectively. In the limit where $\gamma$ and $\delta$ are zero, Eq. (1) is reduced to the fully integrable AL equation that is known to display $M$ conservation laws. In Eq. (1), $f_n$ represents an on-site random normal perturbation in the corresponding channel's propagation constant that is obtained from a Gaussian distribution with zero mean and a root-mean-square width $\sigma = 1$. Given that the perturbation terms in this equation break the lattice symmetries and hence its integrability, one will expect that some of the invariants will vanish. Nonetheless, these perturbations conserve the time and phase shift symmetry, indicating that at least two invariant quantities shall persist. As we will see, in this case, the optical power (norm) and internal energy will still remain 'quasi-invariant'. These two quantities are associated with two conservation laws associated with the integrable AL model, which are given by [32]

$$K^{(1)} = \sum_{n=1}^{M} \ln(1 + |a_n|^2), \quad (2a)$$

$$K^{(2)} = -\sum_{n=1}^{M} (a_n a_{n+1}^* + a_n^* a_{n+1}). \quad (2b)$$

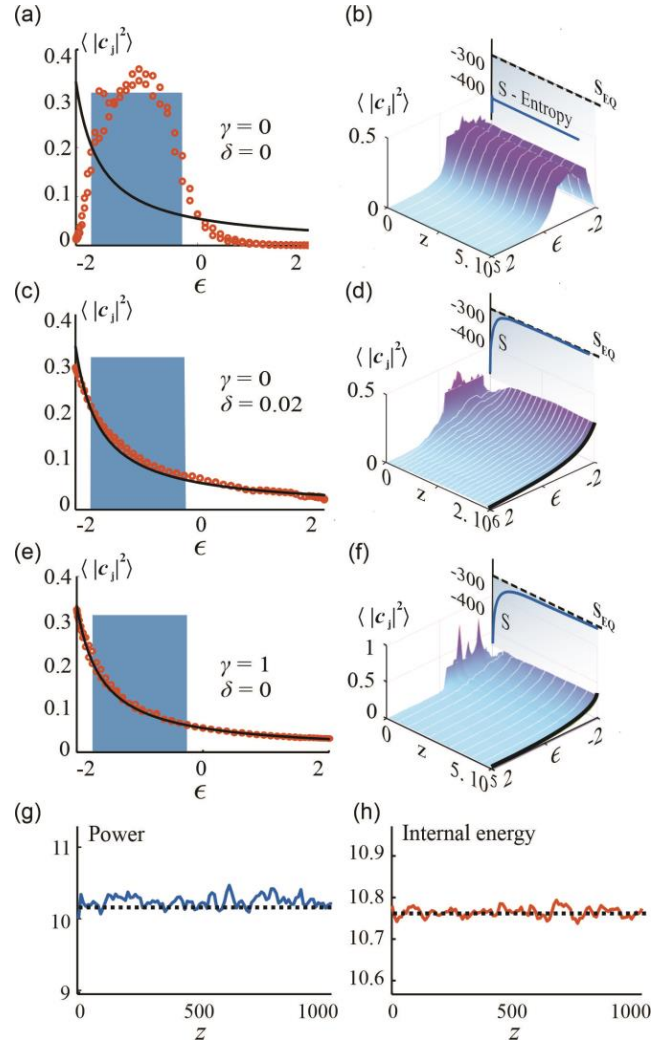

**Fig. 2.** Nonlinear dynamics in unperturbed and perturbed AL lattices when $P = 10$, $U = 10.77$ and $M = 100$. In all cases, the initial power distribution at the input is shaded in blue while the output modal occupancies are depicted by red dots. The black curves represent the anticipated RJ distribution, if occurred. The figures in the right panels show the evolution of the modal occupancies with distance. In (a) and (b) the unperturbed AL system fails to thermalize while in (c)-(f) it settles into a RJ distribution once a small disorder ((c)-(d)) or a Kerr nonlinearity ((e)-(f)) is introduced, respectively. Figures (g) and (h) indicate that $U$ and $P$ remain quasi-invariant during propagation.

In what follows, we will investigate the thermalization dynamics of perturbed AL arrangements under weak nonlinear conditions. Note that in the weakly nonlinear regime, the $K^{(1)}$ invariant is approximately equal to the total power $P$ flowing in the lattice since $ln(1 + x) \approx x$ when $|a_n|^2 \ll 1$. Therefore,

$$P = \sum_{n=1}^{M} |a_n|^2 \approx K^{(1)}. \quad (3)$$

Of importance, will be to express these two constants of motion within the supermode basis of this periodic array. It is worth emphasizing that the invariants of Eqs. (2,3) are for the case of $\gamma = \delta = 0$. In general, the local fields $a_n$ can be obtained from the state vector $|\Psi\rangle$ which in turn can be expressed through the supermodes $|j\rangle$ of the system via $|\Psi\rangle = \sum_{j=1}^{M} c_j |j\rangle \exp(i\epsilon_j z)$ where $c_j$ stand for the complex supermode amplitudes and $\epsilon_j = 2\cos(2\pi j/M)$ represents the eigenvalue of $|j\rangle$. Hence in the supermode basis, the invariants of Eqs. (2,3) can now be expressed as

$$P = \sum_{j=1}^{M} |c_j|^2. \quad \text{(4a)}$$

$$U = -\sum_{n=1}^{M}(a_n a_{n+1}^* + a_n^* a_{n+1}) = -\sum_{j=1}^{M} \epsilon_j |c_j|^2. \quad \text{(4b)}$$

where in establishing Eq. (4a) weakly nonlinear conditions were assumed. The linear Hamiltonian $U$ in Eq. (4b) will now assume the role of the internal energy of the system [19,33]. However, as already mentioned above, these additional $M - 2$ quantities tend to "evaporate", i.e., they will no longer remain constants of motion once the AL system is perturbed and therefore will be inconsequential to the thermalization process. In other words, since only two symmetries are preserved, the system will support only the two quasi-invariants of Eqs. (4). As shown in [19–21], in the presence of these two invariants, the expectation values of the modal occupancies $\langle |c_j|^2 \rangle$ are expected to settle into a Rayleigh-Jeans distribution [19,34–37] once thermal equilibrium is attained, i.e.,

$$\langle |c_j|^2 \rangle = -\frac{T}{\epsilon_j + \mu}. \quad \text{(5)}$$

In Eq. (5), $T$ represents the optical temperature of the photonic lattice system and $\mu$ denotes its corresponding chemical potential. In general, these two intensive variables can be uniquely determined from initial conditions [20]. For demonstration purposes, let us consider an AL array with 100 sites, i.e., $M = 100$. As a first scenario, we study the fully integrable ($\gamma = \delta = 0$) AL lattice when excited with an input power $P = 10$. The initial modal distribution used is extended uniformly within the eigenvalue range $-1.75 \leq \epsilon \leq -0.25$, as shown in Fig. 2(a). The corresponding evolution of the modal occupancies, as obtained after time-averaging in the supermode basis, is depicted in Fig. 2(b). Evidently, in this case, the integrable AL system fails to thermalize and as a result the expectation values of the modal occupancies do not obey the Rayleigh-Jeans (RJ) law. To some extent, this should have been anticipated given that the presence of $M$ invariants does not allow the system to explore the entirety of its phase space.

At this point, one may pose the following question. Can the AL nonlinearity itself, which is nonlocal and non-Hermitian, drive a perturbed AL system into a statistical RJ equilibrium, even though no multi-wave mixing process is engaged [38] and the output is partially coherent [39,40]? Interestingly, even in this case ($\gamma = 0$), the perturbed AL lattice thermalizes into a RJ distribution, as long as an on-site Anderson-like potential ($\delta f_n$) is randomly imposed in the lattice [29,30]. Figures 2(c) and 2(d) depict this scenario when the AL configuration is excited under the same initial conditions as in Figs. 2(a) and 2(b). In these two figures, the linear disorder was set $\delta = 0.02$. In general, one finds that in the weakly nonlinear regime, this non-Hermitian nonlinearity is perfectly capable of thermalizing the AL lattice into a RJ distribution with a definite optical temperature and chemical potential. If the imposed perturbations are relatively weak, then the equilibrium optical temperature $T$ can be directly obtained from the initial conditions $(U, P)$ via $T = (U^2 - 4P^2)/(2UM)$ [19,22]. Once the temperature is known, the chemical potential can be deduced from the equation of state $\mu = (U - MT)/P$ [19,20]. These theoretical results are in good agreement with those obtained from the numerical simulations displayed in Figs. 2(c) and 2(d) where $\mu = 2.4$ and $T = -0.13$. We note that this is unlike what happens in the unperturbed case where no such equilibrium state is reached. From these results, one can conclude that the modal occupancies evolve in a chaotic manner, indicating that the modes are interacting more freely in comparison to the unperturbed case of Figs. 2(a) and 2(b). More importantly, the simulations of Figs. 2(c) and 2(d) totally dispel the commonly held belief that the RJ distribution is merely a byproduct of wave-mixing interactions. Instead, the observed behavior is in accord with a statistical interpretation of the Rayleigh-Jeans distribution when two quasi-conserved quantities ($U$, $P$) are involved [19,21]. Next, we investigate this system when Kerr nonlinearity ($\gamma = 1$) is introduced in the AL periodic lattice ($P = 10, M = 100, \delta = 0$) under the same excitation conditions. Even in this case, the AL system thermalizes relatively fast into a RJ distribution again with $\mu = 2.4$, $T = -0.13$, as shown in Figs. 2(e) and 2(f). As clearly indicated in Figs. 2(d) and 2(f) the optical entropy $S$ [19]

$$S = \sum_{j=1}^{M} \ln|c_j|^2, \quad \text{(6)}$$

always monotonically increases until it maximizes itself in full accord with the second law of thermodynamics. On the other hand, if the system is fully integrable (Fig. 2(b)), the entropy does not attain its maximum value under the imposed initial conditions. The rate of thermalization as a function of normalized coupling lengths can also be inferred from Figs. 2(d) and 2(f). Finally, Figs. 2(g) and 2(h) depict the evolution of $U$ and $P$ for the case corresponding to Figs. 2(c) and 2(d). Indeed, these two quantities remain quasi-invariant as required for thermalization.

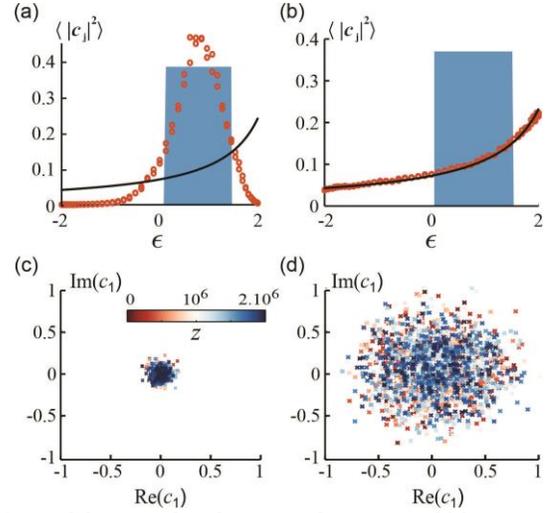

**Fig. 3.** Modal power distribution at the output of (a) an integrable AL system and (b) an AL lattice when a single site ($n = 50$) is perturbed. (c), (d) The phase space for the fundamental mode complex amplitude $c_1$ for a single realization corresponding to (a) and (b).

Next, we examine in greater detail how the system evolves in the two extreme limits, integrable versus non-integrable, by monitoring the phase space of the complex modal amplitudes. In the non-integrable regime, weak disorder ($\delta = 0.1$) is introduced at a particular site ($n = 50$). In both cases, $P = 10$ and $M = 100$ and the same excitation conditions are used in Figs 3(a) and (b). Again, the perfect AL system fails to thermalize (Fig. 3(a)) while the perturbed system settles into a RJ distribution with $\mu = -2.8$, $T = 0.2$. Meanwhile, Figs. 3(c) and 3(d) depict the evolution of the phase space associated with the complex modal amplitude of the ground state $c_1$. Evidently, when the AL lattice is perturbed and is therefore non-integrable, the phase space is considerably enlarged. This is a direct indication that chaos is at play-a necessary ingredient for thermalization.

In conclusion, we have investigated the thermalization dynamics of perfect and modified Ablowitz-Ladik weakly nonlinear optical lattices. We have shown that by introducing a very weak perturbation or local nonlinearity, the once integrable AL lattice is fully capable to thermalize into a Rayleigh-Jeans distribution with a well-defined temperature and chemical potential. Our results clearly indicate that even when a non-local and non-Hermitian nonlinearity is involved the AL system will eventually thermalize in the presence of two quasi-conserved quantities.

**Funding.** This work was partially supported by the Office of Naval Research (ONR) (MURI: N00014-20-1-2789, N00014-18-1-2347, N00014-19-1-2052, N00014-20-1-2522), Air Force Office of Scientific Research (AFOSR) (MURI: FA9550-20-1-0322, MURI: FA9550-21-1-0202), National Science Foundation (NSF) (EECS-1711230, DMR-1420620, ECCS CBET 1805200, ECCS 2000538, ECCS 2011171), US Air Force Research Laboratory (AFRL) (FA86511820019), DARPA (D18AP00058), Army Research Office (W911NF-17-1-0481), Qatar National Research Fund (QNRF) (NPRP13S0121-200126), MPS Simons collaboration (Simons Grant No. 733682), W. M. Keck Foundation, US-Israel Binational Science Foundation (BSF: 2016381). G. G. Pyrialakos acknowledges the support of the Bodossaki Foundation.

## REFERENCES

1. D. N. Christodoulides and R. I. Joseph, Opt. Lett. **13**, Issue 9, pp. 794-796 13, 794 (1988).
2. H. S. Eisenberg, Y. Silberberg, R. Morandotti, A. R. Boyd, and J. S. Aitchison, Phys. Rev. Lett. **81**, 3383 (1998).
3. F. Lederer, G. I. Stegeman, D. N. Christodoulides, G. Assanto, M. Segev, and Y. Silberberg, Phys. Rep. **463**, 1 (2008).
4. J. Fleischer, M. Segev, N. Efremidis, and D. N. Christodoulides, Nature **422**, 147 (2003).
5. R. Morandotti, U. Peschel, J. S. Aitchison, H. S. Eisenberg, and Y. Silberberg, Phys. Rev. Lett. **83**, 4756 (1999).
6. T. Pertsch, F. Lederer, and U. Peschel, Opt. Lett. **23**, Issue 21, pp. 1701-1703 23, 1701 (1998).
7. I. L. Garanovich, A. Szameit, A. A. Sukhorukov, T. Pertsch, W. Krolikowski, S. Nolte, D. Neshev, A. Tünnermann, and Y. S. Kivshar, Opt. Express **15**, 9737 (2007).
8. H. S. Eisenberg, Y. Silberberg, R. Morandotti, and J. S. Aitchison, Phys. Rev. Lett. **85**, 1863 (2000).
9. D. H. Dunlap and V. M. Kenkre, Phys. Rev. B **34**, 3625 (1986).
10. K. Shandarova, C. E. Rüter, D. Kip, K. G. Makris, D. N. Christodoulides, O. Peleg, and M. Segev, Phys. Rev. Lett. **102**, 123905 (2009).
11. M. J. Ablowitz and J. F. Ladik, J. Math Phys. **17**, 1011 (1976).
12. M. J. Ablowitz, M. A. Ablowitz and P. A. Clarkson, *Solitons, Nonlinear Evolution Equations and Inverse Scattering* (Cambridge University Press 1991).
13. A. Szameit, F. Dreisow, M. Heinrich, S. Nolte, and A. A. Sukhorukov, Phys. Rev. Lett. **106**, 1 (2011).
14. X. Y. Wu, B. Tian, L. Liu, and Y. Sun, Communications in Nonlinear Science and Numerical Simulation **50**, 201 (2017).
15. P. G. Kevrekidis, G. J. Herring, S. Lafortune, and Q. E. Hoq, Phys. Lett. A **376**, 982 (2012).
16. X. Y. Xie, B. Tian, X. Y. Wu, and Y. Jiang, Optical Engineering **55**, 106122 (2016).
17. H. N. Chan and K. W. Chow, Comm. Nonlinear Science Numerical Simulation **65**, 185 (2018).
18. U. Al Khawaja and A. A. Sukhorukov, Opt. Lett. **40**, 2719 (2015).
19. F. O. Wu, A. U. Hassan, and D. N. Christodoulides, Nat. Photonics **13**, 776 (2019).
20. M. Parto, F. O. Wu, P. S. Jung, K. Makris, and D. N. Christodoulides, Opt. Lett. **44**, 3936 (2019).
21. K. G. Makris, Fan. O. Wu, P. S. Jung, and D. N. Christodoulides, Opt. Lett. **45**, 1651 (2020).
22. F. O. Wu, P. S. Jung, M. Parto, M. Khajavikhan, and D. N. Christodoulides, Nat. Communications Physics **3**, (2020).
23. C. Shi, T. Kottos, and B. Shapiro, Physical Review Research **3**, 033219 (2021).
24. A. Ramos, L. Fernández-Alcázar, T. Kottos, and B. Shapiro, Phys. Rev. X **10**, 031024 (2020).
25. N. K. Efremidis and D. N. Christodoulides, Phys Rev A (Coll Park) **103**, 043517 (2021).
26. G. G. Pyrialakos, H. Ren, P. S. Jung, M. Khajavikhan, and D. N. Christodoulides, Phys. Rev. Lett. 128, 213901 (2022).
27. P. S. Jung, G. G. Pyrialakos, F. O. Wu, M. Parto, M. Khajavikhan, W. Krolikowski, and D. N. Christodoulides, Nat. Comm. 13, 4393 (2022).
28. M. Baldovin, A. Vulpiani, and G. Gradenigo, Journal Statistical Physics **183**, (2021).
29. M. Segev, Y. Silberberg, and D. N. Christodoulides, Nat. Photonics **7**, 197 (2013).
30. Y. Lahini, A. Avidan, F. Pozzi, M. Sorel, R. Morandotti, D. N. Christodoulides, and Y. Silberberg, Phys. Rev. Lett. **100**, 013906 (2008).
31. T. Schwartz, G. Bartal, S. Fishman, and M. Segev, Nature **446**, 52-55 (2007).
32. M. J. Ablowitz, B. Prinari, and A. D. Trubatch, *Discrete and continuous nonlinear Schrödinger systems* (Cambridge University Press, 2004).
33. M. A. Selim, F. O. Wu, H. Ren, M. Khajavikhan, and D. Christodoulides, Phys. Rev. A **105**, (2022).
34. H. Pourbeyram, P. Sidorenko, F. O. Wu, N. Bender, L. Wright, D. N. Christodoulides, and F. Wise, Nature Physics **18**, 1-6, (2022).
35. E. V. Podivilov, F. Mangini, O. S. Sidelnikov, M. Ferraro, M. Gervaziev, D. S. Kharenko, M. Zitelli, M. P. Fedoruk, S. A. Babin, and S. Wabnitz, Phys. Rev. Lett. **128**, 243901 (2022).
36. F. Mangini, M. Gervaziev, M. Ferraro, D. S. Kharenko, M. Zitelli, Y. Sun, V. Couderc, E. v. Podivilov, S. A. Babin, and S. Wabnitz, Opt. Express, Vol. **30**, Issue 7, pp. 10850-10865 30, 10850 (2022).
37. F. Mangini, M. Ferraro, M. Gervaziev, D. S. Kharenko, M. Zitelli, Y. Sun, V. Couderc, E. V. Podivilov, S. A. Babin, and S. Wabnitz, Optica Advanced Photonics Congress, (2022).
38. S. Dyachenko, A. C. Newell, A. Pushkarev, and V. E. Zakharov, Physica D **57**, 96 (1992).
39. M. A. Selim, F. O. Wu, G. G. Pyrialakos, M. Khajavikhan, D. Christodoulides, arXiv preprint arXiv:2212.10063 (2022).
40. M. A. Selim, F. O. Wu, G. G. Pyrialakos, M. Khajavikhan, D. Christodoulides, CLEO: QELS_Fundamental Science. Optica Publishing Group, 2022.